\documentclass{aastex}
\begin{document}

\title{An HST STIS Observation of VW Hydri at the Exact FUV Onset of an
Outburst \footnote{Based on observations made with the {\it{Hubble Space 
Telescope}}. {\it{HST}} is
operated by the Association of Universities for Research in Astronomy,
Inc., under NASA Contract NAS5-26555}} 
 
\author{Edward M. Sion}
\affil{Department of Astronomy and Astrophysics, Villanova University, \\ 
800 Lancaster Avenue, Villanova, PA 19085, USA}
\email{edward.sion@villanova.edu}

\author{F. H. Cheng}
\affil{Department of Physics,
Shanghai University, 99 Shang-Da Road,
Shanghai 200436, P.R. China}
\email{cheng3@prodigy.net}

\author{Boris T. G\"ansicke}
\affil{Department of Physics, University of Warwick
Coventry CV4 7AL, United Kingdom}
\email{Boris.Gaensicke@warwick.ac.uk}

\author{Paula Szkody}
\affil{Department of Astronomy,
University of Washington,
Seattle, WA 98195, USA}
\email{szkody@alicar.astro.washington.edu}
\clearpage

\begin{abstract} 
We present an analysis of HST STIS data of VW Hyi that we
acquired ~ 14 days after a superoutburst. At the time of our observation
the system appears to be going into outburst with the longest wavelengths
increasing in flux by a factor of 5 while the shortest wavelengths
increase by only a factor of 2. Using the distance of 65 pc, a system
inclination angle of 60 degrees and a white dwarf mass of 0.86
$M_{\odot}$, we carried out model fits involving a white dwarf by itself,
an optically accretion disk by itself, a composite model using an
optically thick accretion disk and a white dwarf, a two-temperature white
dwarf model with a cooler more slowly rotating photosphere and a hotter,
rapidly rotating accretion belt and a composite model involving a white
dwarf and a rapidly rotating cooler disk ring heated up to 
a ``low'' temperature of 
~13-14,000K. This component of temperature stays fairly constant throughout the HST 
observations while the area of the disk ring increases by a factor of 12. 
We see evidence of a delay in the UV emission consistent with the outburst beginning
outside of a disk truncation radius. 
\end{abstract}

\keywords{Stars: Accretion, Dwarf Novae, White Dwarfs, Chemical
Abundances, Individual: VW Hydri}

\section{Introduction}

The dwarf nova VW Hydri is a well-studied SU UMa-type system having both
normal and superoutbursts and lying below the cataclysmic variable period
gap with P$_{orb} = 0.074$ days \citep{war95}. 
It is one of several cataclysmic variables in which the
underlying white dwarf accreter is exposed during dwarf nova quiescence.  
These systems offer the prospect of detailed far UV spectroscopic analyzes
which have yielded rotation rates, surface temperatures, dynamical masses,
and chemical abundances of their accreted atmospheres 
\citep{sio99,gan99}.  Previous studies of the white dwarf in VW Hyi
revealed its white dwarf has T$_{eff} = 19,000$K \citep{sio95a,sio95b},
mass M$_{wd} = 0.86$ M$_{\sun}$ \citep{sio97}, a nitrogen to carbon
abundance of 4-5 times the solar ratio and early indications of
nucleosynthetic abundances from proton-capture reactions (including P, Al,
and Mn) greatly elevated above solar in its accreted atmosphere 
\citep{sio97,sio01} and a rapidly spinning region hotter than the
T$_{eff}$ of the white dwarf \citep{sio96,gan96}.  All of the above
characteristics were confirmed by the HST Cycle 8 STIS abundance study of
\citet{sio01}. As part of this study of the white dwarf during quiescence
to get a radial velocity curve, our observation fortuitously covered the
beginning of the transition into a normal outburst of VW Hydri. In this
paper, we report the results of our analysis of this unforeseen event and
its implications for understanding the observed behavior of VW Hydri.

\section{Observations}

The observations of VW Hydri were obtained on December 10, 2001, starting
at 20:28:00 UT, $\sim 14$ days after the return to optical quiescence
following the November 2001 superoutburst. During the STIS observation of
VW Hyi, we serendipitously recorded the onset of a normal outburst (up to
$\sim 12^{th}$ magnitude) which eventually reached V = 10.1.  
The instrumental setup used STIS with the
FUV-MAMA detectors in ACCUM mode and the $0.2'' \times 0.2''$ aperture.
The disperser was the E140M with a center wavelength of 1425\AA, a
wavelength range of 1140.0 \AA\ to 1735 \AA\ and a total exposure of
9120 seconds. The exposure was subdivided into 19 individual subexposures
which accurately record the remarkable changes in the spectrum. The
evolution of the spectra is displayed in figure 1. The first seven spectra
looked quite similar with roughly the same flux level. The changes began
to be noticeable in spectrum 8 when the flux level rose slightly and
emission peaks began to appear. These emission regions showed a marked
departure from the flux predicted by a single temperature white dwarf
model.

The dramatic change during the observation can also be characterized
by examining the flux ratios at the short end, middle and long wavelength
end of the STIS spectra. To see the largest change, the fluxes at 1160A,
1450A and 1725A are compared for spectra 1 and 19. The flux ratio of 
spectrum 1 to spectrum 19 at 1160A is only 1.75 while the flux changes at
1450A and 1725A are 5.0 and 4.7 respectively.

\begin{deluxetable}{cccccccc}
\tablewidth{0pc}
\tablecaption{FUV Flux Variation} 
\tablehead{ 
Obs.No. & Flux (1160)           & Flux (1450) & Flux (1725) & Ratio & 
1160  & 1450  & 1725 \\  
  &$<$erg/s/cm$^2$/\AA $>$& $<$erg/s/cm$^2$/\AA $>$ &$<$erg/s/cm$^2$/\AA $>$&  
 19/1 &     &       &      
} 
\startdata 
1 & $2.0\times 10^{-13}$&$2.0\times 10^{-13}$&$1.5\times 10^{-13}$& & & & \\ 
19 & $3.5\times 10^{-13}$&$1.0\times 10^{-12}$&$7.0\times 10^{-13}$&&1.8&5.0&4.7 \\ 
\enddata
\end{deluxetable} 

This flux behavior suggests the possibility we are seeing the so-called UV
delay where the UV flux lags behind the optical by as much as 12 hours or
more during the rise to outburst of many dwarf novae \citep{liv92}.
Unfortunately, there is insufficient time resolution in the optical
observations at the time of our HST observation to construct an optical
light curve for comparison with the FUV light curve. Two days prior to the
HST observation, VW Hydri was fainter than V = 13.5 while 1 day after the
HST observation, VW Hyi had risen to V = 10.1, according to the AAVSO
archive.

Overall, the line spectrum in this observation has the same strong
absorption features and mix of ions as earlier HST FOS and GHRS spectra
\citep{sio95a,sio95b,sio96,sio97} of VW Hydri after outburst or
superoutburst (e.g. Si II (1260, 1265), Si III (1300), C II (1335), Si IV
(1393, 1402), Si II (1526, 1533), with C IV (1548, 1550) in moderately
strong emission). The only emission features in the spectra are C IV
(1548, 1550) during the early spectra and a feature near 1350A (C II
1335?) which peaks up as the outburst progresses.

\subsection{Synthetic Spectral Fitting}

The model atmosphere \citep[-TLUSTY]{hub88}, and spectrum synthesis
\citep[-SYNSPEC]{hub94,hub95} codes and details of our $\chi^{2}$
minimization fitting procedures can be found in \citet{sio95a,sio95b} and
\citet{hua96} and for the sake of 
brevity will not be repeated here. We fixed the distance
of VW Hyi at d = 65 pc, fixed the white dwarf radius at R$_{wd} =
9.8\times 10^{-3} R_{\sun}$, corresponding to the gravitational redshift
mass M$_{wd} = 0.86 M_{\sun}$ \citep{sio97}. Taking the fixed gravity
log $g = 8.5$ we created white dwarf models for temperatures $T/1000 =
14 - 23.5$ K in steps of 0.5, rotational velocities V$_{rot} = 200 - 700$
km/s in steps of 100 km/s and chemical abundances Si = 0.1, 0.2, 0.3, 0.4,
0.5 x solar abundance, and C = 0.3, 0.4, 0.5, 0.6, 0.7 x solar abundance
with all other elements at their solar values initially. This initial grid
comprised a total of 3000 models. The white dwarf models have been
normalized to a solar radius and a distance of 1 kpc. The model fluxes are
scaled to the observed fluxes through the scale factor S$ = 4 \times \pi
(R_{wd}^{2}/D^{2}$).  For the disk models, we made use of the entire
parameter space covered by the optically thick accretion disk model grid of
\citet{wad98} for $\dot{M}$ but held the disk inclination and the white dwarf
mass fixed as above. Our accretion belt models were constructed in the
same manner as described in \citet{sio96}.  A detailed discussion of the
accretion belt models can also be found in \citet{sio01}.

First, we applied the single temperature white dwarf models to the data.
The temperature remained roughly constant (within the uncertainties of T)  
through spectrum 8 but subsequent observed spectra revealed large changes in the
flux level longward of 1600A and relatively small changes in the flux
level at shorter wavelengths down to around Lyman Alpha. None of the model
fits are successful in a statistical or a qualitative sense.  Therefore,
the term ``best-fitting'' refers to the lowest $\chi^2$ value found for each
of the 19 spectra. The ``best-fitting'' white dwarf models to spectra 1 and
19 are displayed in figure 2 and 3. At the beginning of the exposure
T$_{eff}$ = 24,000K, $V\sin i$ = 600 km/s with $\chi^{2}$ = 15.1767 while
for spectrum 19, T$_{eff}$ = 32,000K, $V\sin i$ = 600 km/s with $\chi^{2}$
= 445.953.

Second, we tried optically thick accretion disk models alone without any
white dwarf photospheric radiation contribution. There is very little
qualitative improvement apparent and the disk fits, like the photosphere
fits, agree very poorly with the observations. Likewise for
two-temperature white dwarf models (photosphere plus hotter equatorial
belt), the fits are similarly unsatisfactory as they do not account for
the long wavelength flux increase. The size and variation of the $\chi^2$
values are very similar to those obtained for the single-temperature
photosphere fits.

Clearly, none of the above fitting experiments provided agreement, either
qualitatively or quantitatively, with the STIS observations. In view of
this failure and the fact that the longward continuum rises while the
Lyman Alpha region remains relatively unchanged, we explored the
possibility of a cool accretion ring existing around the white dwarf.

We created a grid of white dwarf plus cool accretion disk ring models with the
white dwarf temperature, disk ring temperature, disk ring gravity, disk ring velocity,
and disk ring area as free parameters. The system distance, white dwarf gravity
and rotational velocity, were kept fixed. The chemical abundances of the
white dwarf were also kept fixed at the solar values. We let the white
dwarf temperature vary between 18 and 28,000K in steps of 1000K. The trial
temperatures of the ring were 12,000K, 13,000K, 14,000K and 15,000K, the
trial ring velocities were 3400 km/s, 2500 km/s, 1500 km/s and 1000 km/s,
the chemical abundance of the ring was kept fixed at the solar values. The
ratio of the surface area of the ring to the surface area of the white
dwarf was varied from 0.5 to 100 in steps of 0.5.

When a white dwarf model is combined with a cooler accretion ring, this results in a
very significant improvement in the quality of the fits. In the spectrum
of the undisturbed white dwarf, the flux at 1160A is normally higher than
the flux at 1725A. Starting approximately with spectrum 5, the flux at
1725A first starts to surpass the flux level at the shortest wavelength,
1160A. To illustrate the improvement in fitting the evolution of the
spectra, it is illuminating to compare spectrum 5 with spectrum 19. In
figure 4, the best-fitting composite ring model to spectrum 5 is
displayed. For comparison, the best-fitting composite ring model is
displayed for spectrum 19 in figure 5. While the fits are far from
perfect, they do reproduce the overall spectral shape recorded with STIS.
The only discrepancy remaining is that the continuum seems lower than that
observed in the 1450-1600A region and some of the absorption lines are not
well fit in that region. From the fitting of the evolution of the spectra,
we see that the white dwarf temperature appears to increase from 21,000k
to 27,000K. This temperature change is of the same order as the change in
temperature of the white dwarf in U Gem in response to its outburst. The
temperature of the rotating ring changed only insignificantly, from 14,000
to 13,000K, however the area of the rotating ring, relative to the area of
the white dwarf, increased by a factor of ~ 12!

\section{Conclusions}

The rapid flux increase during the course of the observation starting with
spectrum 5 and reaching a peak flux of $1\times 10^{-12}$ ergs/cm$^2$/s
appears to mark the actual FUV onset of a normal outburst of VW Hyi. The
fact that the spectral energy distribution during the marked increase in
flux undergoes a substantially larger change at longer UV wavelengths
suggests that the outburst began in the outer part of the disk. If the
disk is not truncated, then the shape of the overall flux suggests an
outside-in type outburst. If however the disk is truncated, then all we
can say is that the outburst began external to the truncation radius.

The evolution of the spectrum is surprisingly well-represented by a white
dwarf which is being heated at the time of the observation and a
rapidly rotating accretion ring, cooler than the white dwarf and expanding in
surface area. The greatest change in the STIS spectrum occurs longward of
Lyman Alpha suggesting that the events in the disk have not yet fully
affected the white dwarf.  We have found that the overall evolution of the
distribution of spectral energy is well-represented by a white dwarf and a
cooler accretion ring of increasing area as the outburst proceeds and of
much larger surface area than the white dwarf. We see evidence of a delay
in UV emission consistent with outburst beginning outside of a disk
truncation radius. While the sparse optical coverage versus time during
the HST exposure disallows a detailed comparison with the temporal
variation of the FUV flux, the evolution of the STIS spectra suggest that
the UV emission from the white dwarf is delayed relative to the optical
rise. If the disk is truncated, then this delay is probably caused by the
time it takes viscous diffusion inside the disk truncation radius to fill
in the truncated region with gas and reach the white dwarf.

\acknowledgements
This work is supported by NASA through grants
GO8139.01-97A from the 
Space Telescope Science Institute,
 which is
operated by the Association of Universities for Research in Astronomy,
Inc., under NASA Contract NAS5-26555. 
 Support was also provided in part
by NSF grant 99-01195 and NASA ADP grant NAG5-8388.  BTG was supported by a 
PPARC Advanced Fellowship.

Figure Captions

Figure 1: The sequence of 19 ACCUM spectra from the start to the finish of
the 10 December, 2001 STIS observation. Note the dramatic changes in the
spectrum from the beginning to the end of the exposure.

Figure 2:  The best fit synthetic spectrum of a white dwarf alone, to
spectrum $1, \sim14$ days post-superoutburst.(T$_{eff} = 24,000$K, log $g
= 7.5$, V$_{sin} = 600$ km/s, Si abundance 0.5 solar, C abundance 1.0
solar (see text for details);

Figure 3:  The best fit synthetic spectrum of a white dwarf alone, to
spectrum $19,\sim 14$ days post-superoutburst.(T$_{eff} = 32,000$K, log $g
= 8.5$, V$_{sin} = 600$ km/s, Si abundance 0.2 solar, C abundance 0.1
solar (see text for details);

Figure 4: The best-fitting composite white dwarf plus rotating ring model
to spectrum 5. The white dwarf has $T_{eff}$ = 21,000K and provides 64\%
of the light while the rotating ring has $T _{ring}$ = 14,000K, log g = 6,
$V_{ring}$ = 3400 km/s and provides 36\% of the light. The dotted line is
the white dwarf synthetic flux, the dashed line is the ring synthetic flux
and the solid line is the combined flux of the two components.

Figure 5: The best-fitting composite white dwarf plus rotating ring model
to spectrum 19. The white dwarf now has $T_{eff}$ = 27,000K and provides 36\%
of the light while the rotating ring has $T_{ring}$ = 13,000K, log g = 7,
$V_{ring}$ = 1500 km/s and provides 64\% of the light. The dotted line is the
white dwarf synthetic flux, the dashed line is the ring synthetic flux and
the solid line is the combined flux of the two components.

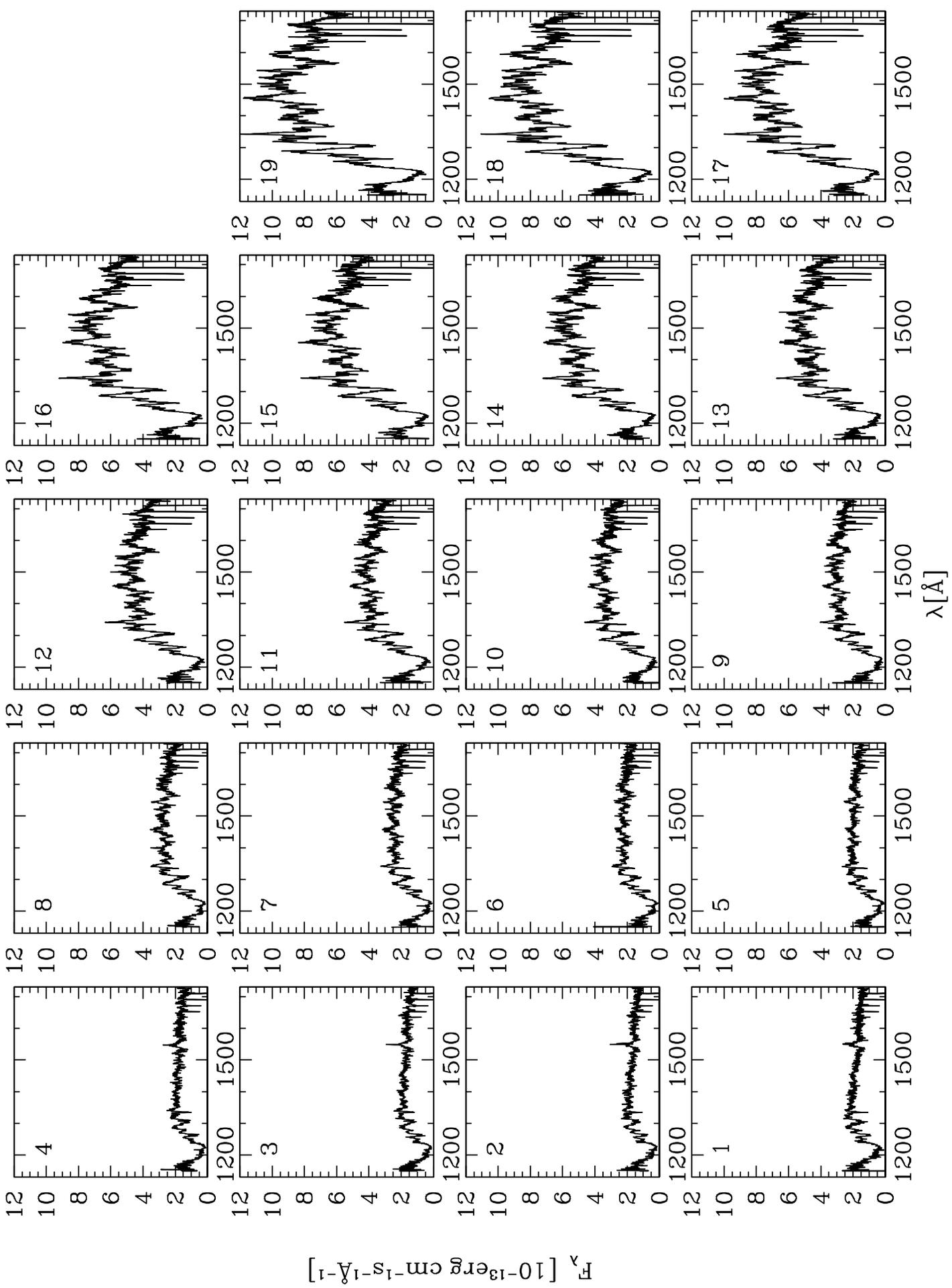
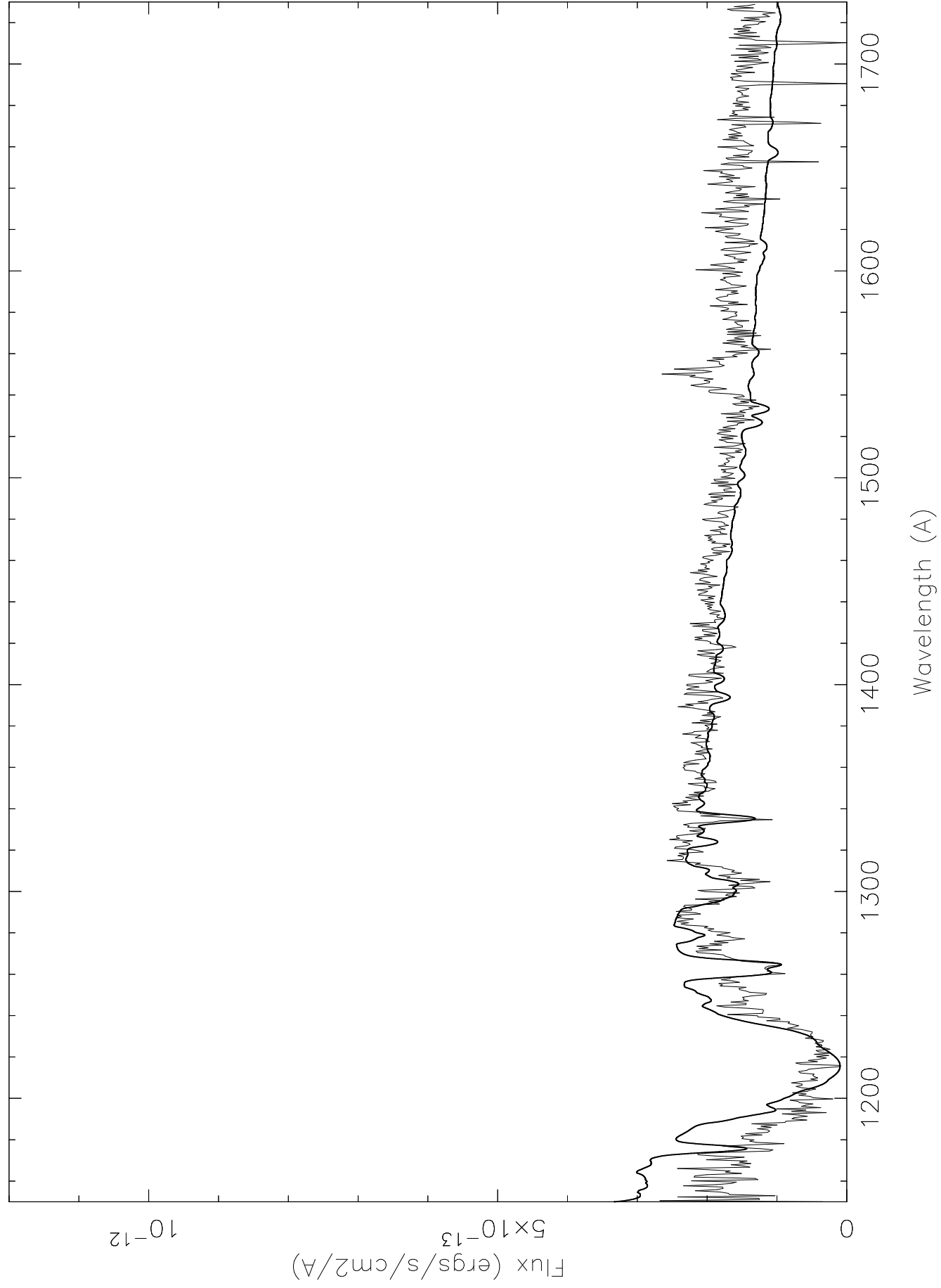
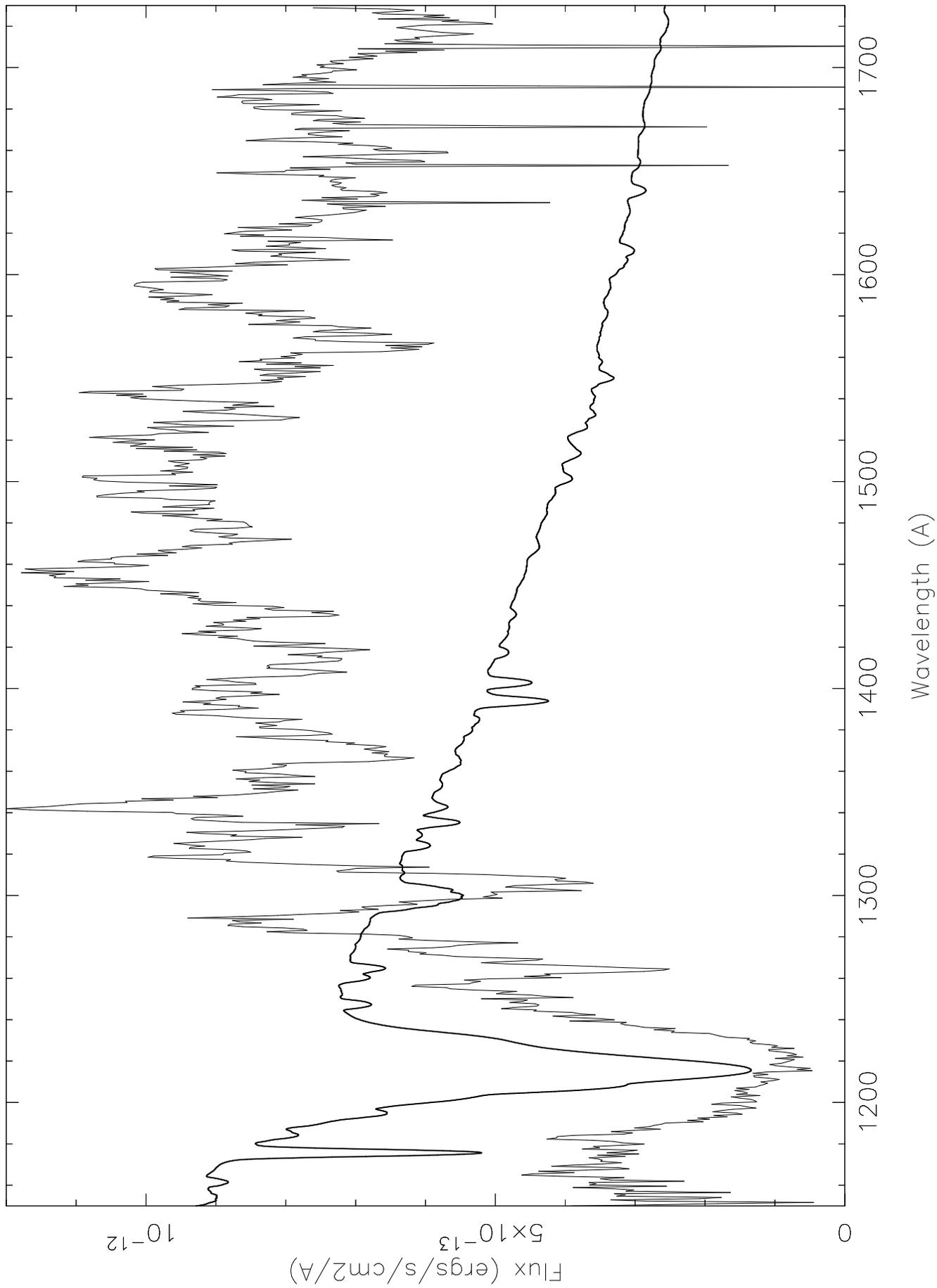
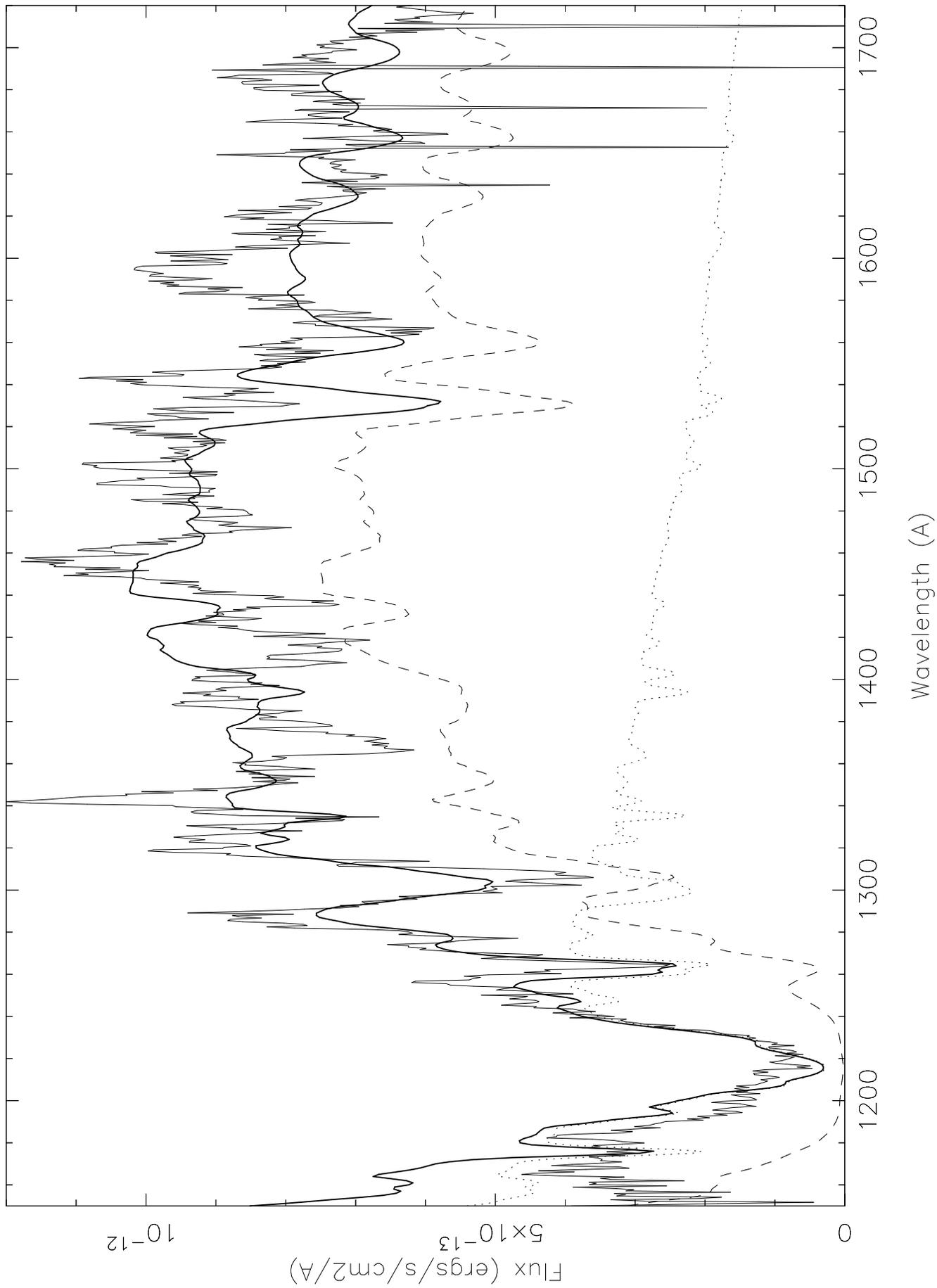
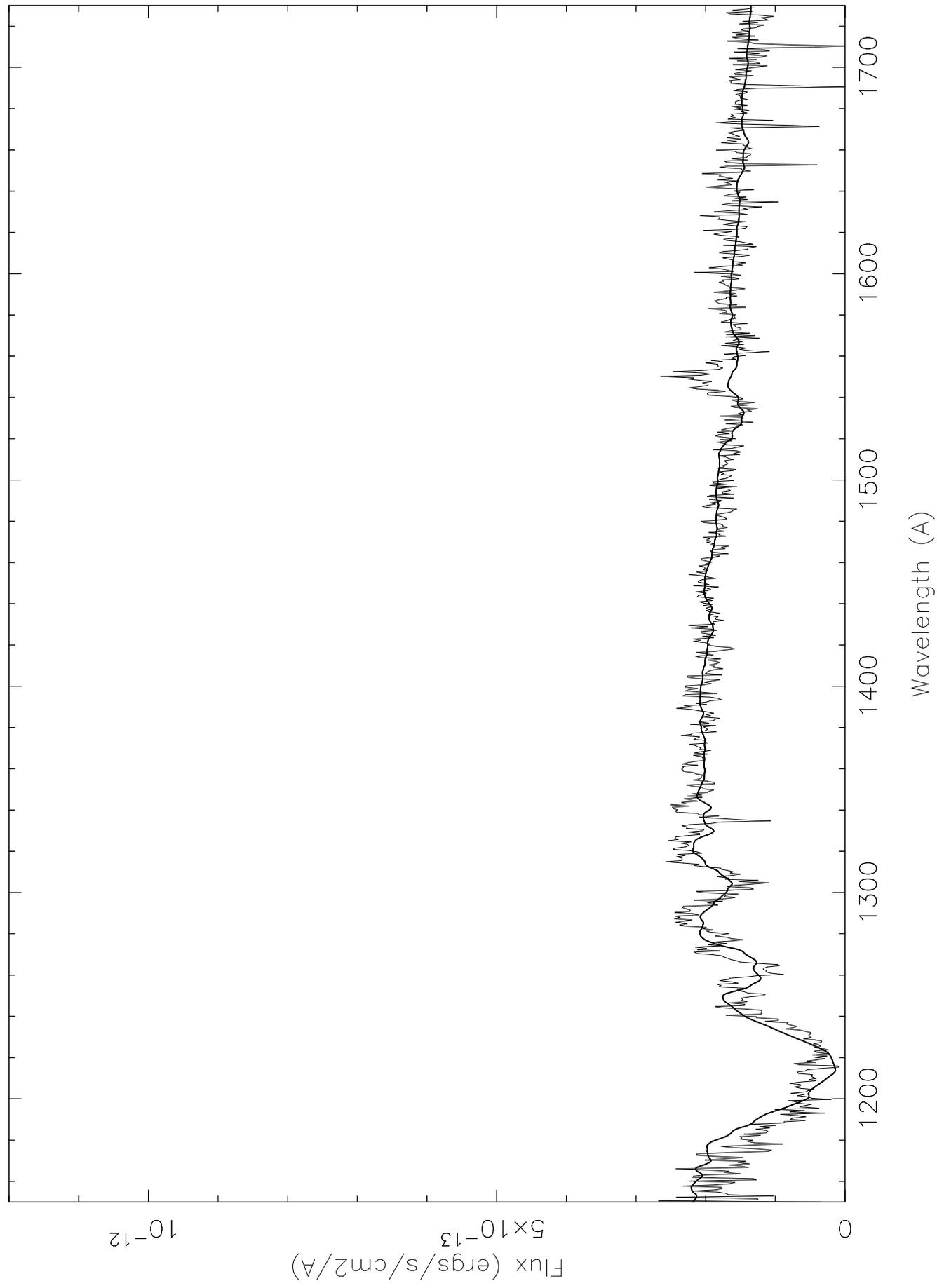
\end{document}